\documentclass[10pt]{article}

\usepackage{amssymb}
\usepackage{amsmath}
\usepackage{graphicx}
\numberwithin{equation}{section}
\def \ccomma{\raise 2pt\hbox{,}\ } 
\def \D {\hbox{d}}
\def \paramb {b}
\def \Npoles {P}
\def \sech{\mathop{\rm sech}\nolimits}
\def \pq{\mathop{\rm pq}\nolimits}
\def \mod#1{\vert #1 \vert}

\def \GLA{A}
\def \csi{\kappa_{\rm i}} 

\begin{document}

\title{Methods for exact solutions of nonlinear ordinary differential equations\\
}
\author{Robert Conte${}^{1,2}$,
        Micheline Musette${}^{3}$,
				Tuen Wai Ng${}^{2}$ and
				Chengfa Wu${}^{4}$
{}\\
\\ 1. Universit\'e Paris-Saclay, ENS Paris-Saclay, CNRS
\\    Centre Borelli, F-91190 Gif-sur-Yvette, France
\\
\\ 2. Department of mathematics, The University of Hong Kong,
\\ Pokfulam, Hong Kong
\\
\\ 3. Dienst Theoretische Natuurkunde, Vrije Universiteit Brussel, 
\\ Pleinlaan 2, B–1050 Brussels, Belgium
\\
\\ 4. Institute for Advanced Study, Shenzhen University, Shenzhen, PR China
\\
\\    E-mail Robert.Conte@cea.fr,         ORCID https://orcid.org/0000-0002-1840-5095
\\    E-mail Micheline.Musette@gmail.com, ORCID https://orcid.org/0000-0002-2442-9579
\\    E-mail NTW@maths.hku.hk,            ORCID https://orcid.org/0000-0002-3985-5132
\\    E-mail CFWu@szu.edu.cn,             ORCID https://orcid.org/0000-0003-1697-4654
{}\\
}

\maketitle

\hfill

{\vglue -10.0 truemm}



\begin{abstract}
In order to find closed form solutions of nonintegrable nonlinear ordinary differential equations,
numerous tricks have been proposed.
The goal of this short review is to recall classical, 19th-century results,
completed in 2006 by Eremenko,
which can be turned into algorithms, thus avoiding \textit{ad hoc} assumptions,
able to provide \textit{all} (as opposed to some) solutions
in a precise class.
To illustrate these methods,
we present some new such exact solutions, physically relevent. 
\end{abstract}

\noindent \textit{Keywords}:

\noindent \textit{PACS 1995}~:
 02.30.-f,  
 02.70.-c,  
 05.45.+b,  
 47.27.-i.  

\baselineskip=12truept


\tableofcontents

\section{Introduction. Sufficient \textit{vs}.~necessary, tricks \textit{vs}.~methods}

The question addressed in this review is the following.
Given some nonlinear algebraic autonomous ordinary differential equation (ODE),
to find as many inequivalent, closed form solutions as possible.
Let us first define this vocabulory.

The ODE is assumed to be polynomial in all the derivatives of the function $u(x)$ (``algebraic''),
with constant coefficients (``autonomous'').
The precision ``inequivalent'' is important. 
For instance, given the ODE
\begin{eqnarray}
& &  \frac{\D u}{\D x} + u^2-1=0,
\nonumber
\end{eqnarray}
the three expressions
\begin{eqnarray}
& &  \tanh(x-x_1),  \coth(x-x_2), \frac{\tanh(x)-c}{1+ c \tanh(x)}\ccomma (x_1,x_2,c) \hbox{ constants},
\nonumber
\end{eqnarray}
are equivalent because exchanged by a translation of $x$ 
(a consequence of the addition formula of trigonometric functions),
therefore presenting them as different is incorrect.
Similarly, given the ODE
\begin{eqnarray}
& &  \left(\frac{\D u}{\D x}\right)^2 = a (u^2-b) (u^2-c),
\label{eqJacobi}
\end{eqnarray}
with $a,b,c$ complex constants,
its general solution can be presented as twelve equivalent expressions 
\begin{eqnarray}
& &  c_1 \pq(k (x-c_2)),
\nonumber 
\end{eqnarray} 
in which the complex constants $c_1,k$ depend on $(a,b,c)$ and $c_2$ is arbitrary,
because of various identities between the Jacobi elliptic functions $\pq$'s 
available in any textbook \cite[Chap.~16, \S 16.8, \S 16.10]{AbramowitzStegun},
therefore one should not list all of them in a publication,
as is sometimes done.
Even worse,
the addition formulae of elliptic functions \cite[Chap.~16, \S 16.17]{AbramowitzStegun}
allows twelve more expressions of the solution of (\ref{eqJacobi}).

\medskip

We will distinguish ``sufficient'' methods from ``necessary'' ones,
and put the emphasis on the second ones.

The \textit{sufficient methods}
assume for the solution a given expression with adjustable coefficients.
By construction, they cannot find solutions outside the given class
(this is the well known story of the drunken man under a lamp post).
For instance, the class of polynomials in $\tanh$ and $\sech$ 
\cite{JeffreyXu,CM1992},
so fruitful to find solutions often observed in physics,
cannot find a solution rational in $\tanh$,
such as the defect solution \cite[Eq.~(9)]{CMNW-CGL35-Letter} 
\begin{eqnarray}
& & 
\mod{\GLA}^2=- 20 \frac{d_i}{e_i} 
 \frac{       \coth^2 \displaystyle\frac{k \xi}{2} 
        \left(\coth^2 \displaystyle\frac{k \xi}{2}-1 \right)}
      {\left(5\coth^2 \displaystyle\frac{k \xi}{2}-3 \right)^2 -12}\ccomma \xi=x-ct,
\nonumber
\end{eqnarray}
($d_i, e_i, k^2, c $ being nonzero real constants)
of the well known quintic complex Ginzburg-Landau equation (CGL5),
\begin{eqnarray}
& &  {\hskip -18.0 truemm}
\hbox{(CGL5)}\
i \GLA_t +p \GLA_{xx} +q \mod{\GLA}^2 \GLA +r \mod{\GLA}^4 \GLA -i \gamma \GLA =0,\
p\ r\ \gamma\ \Im(r/p)\not=0,
\label{eqCGL5}
\end{eqnarray}
in which $p,q,r$ are complex constants and $\gamma$ a real constant.

\medskip

Similarly, the class \cite{Samsonov1994,KudryashovElliptic} of differential polynomials
of either Weierstrass elliptic function $\wp(x)$
or Jacobi elliptic functions $\pq(k x)$,
which has indeed produced many new solutions,
cannot find more general elliptic solutions
like the particular solution of CGL5 found by Vernov
\cite{VernovCGL5},
\begin{eqnarray}
& & 
%
\mod{\GLA}^2=\left(\frac{4 g_r}{3 e_i}\right)^{1/2} \frac{4 \wp^2(\xi,g_2,0)-g_2}{4 \wp^2(\xi,g_2,0)+g_2}\ccomma
g_2=- \frac{g_r^2}{27}\ccomma
{\wp'}^2=4 \wp^3-g_2 \wp-g_3,
\label{eqCGL5-Vernov-solution-M}
\end{eqnarray}
in which $e_i, g_r, g_2, g_3$ are nonzero real constants.

Therefore, general methods are required.

	Innumerable such ``new methods'' are regularly published,
such as the 
``Exp-method'', 
``$G'/G$ expansion method'', 
``simplest equation method'', 
``homogeneous balance method'', 
etc,
but they are only copies of the just mentioned methods
(essentially, class of differential polynomials of either $\wp$ or $\pq$
or their degeneracies $\coth$),
see the criticisms of Refs
\cite{Seven-common-errors}
and
\cite{More-common-errors}. 

As opposed to the above described sufficient methods,
there exist what we will call 
\textit{necessary methods}
able to find \textit{all} (as opposed to some) solutions
in a natural class,
provided the considered ODE possesses two properties very easy to check.

This paper is organized as follows.
In section \ref{sectionExamples},
we first present various equations of physical interest,
to be later processed by the necessary methods.

In section \ref{sectionEremenko},
we recall 
a very nice theorem by Eremenko
which splits autonomous algebraic ODEs in two disjoint subsets:
those ODEs whose \textit{all} meromorphic solutions can be found explicitly,
those for which  \textit{some} (but possibly not all) such solutions can be found.

Section \ref{sectionMethods}
presents constructive methods to implement the theorem by Eremenko.

The next sections 
\ref{section-Application-Quartic},
\ref{section-Application-CGL5},
\ref{section-Application-CGL5plus}
provide various illustrations of these methods.

Finally, we mention in section \ref{section-nonmeromorphic}
an original subequation method, due to Nisha \textit{at alii} \cite{NMGRK2020},
providing nonmeromorphic, multivalued closed form particular solutions.

\section{Our examples: a few equations of physical interest}
\label{sectionExamples}

To illustrate the methods described here,
we choose a few examples taken from physics.

\begin{enumerate}

	\item Our first example 
\cite{NLS-fourth-order-dispersion1994} 
describes a fourth order dispersion in optical fibers \cite[Eq.~(1)]{Parker-Aceves2021},
\begin{eqnarray}
& &  i A_t + \frac{b_4}{24} A_{xxxx} - \frac{b_2}{2} A_{xx} + \gamma |A|^2 A=0,
\nonumber
\end{eqnarray}
with $b_4, b_2, \gamma$ real parameters,
for which the standing wave assumption 
\begin{eqnarray}
& & A(x,t)=u(x) e^{-i \omega t}, u \hbox{ real function},
\nonumber
\end{eqnarray}
generates the ODE
\begin{eqnarray}
\frac{b_4}{24} u_{xxxx} - \frac{b_2}{2} u_{xx} + \gamma u^3 - \omega u =0.
\label{eqQuartic-ODE4}
\end{eqnarray}

	\item 	
	Our second example is the quintic complex Ginzburg-Landau equation (CGL5) (\ref{eqCGL5}).
Its traveling wave reduction
\begin{eqnarray}
& & {\hskip -15.0 truemm}
\GLA(x,t) =\sqrt{M(\xi)} e^{ i(\displaystyle{-\omega t + \varphi(\xi)})},
\xi=x-ct,\
c \hbox{ and }\omega \in \mathbb{R},\
\label{eqCGL35ReducMphi}
\end{eqnarray}
defines, by the elimination of $\varphi$,
an ODE for $M(\xi)$ of third order and second degree \cite[p.~18]{Klyachkin1989} \cite{MC2003},
\begin{eqnarray}
& &
(G'-2 c s_i G)^2 - 4 G M^2  (e_i M^2 + d_i M - g_r)^2=0,\
\label{eqCGL35Order3}
\\ & &
G=\frac{1}{2} M M'' - \frac{1}{4} M'^2
  -\frac{c s_i}{2} M M'  + g_i M^2 + d_r M^3 + e_r M^4,
\nonumber
\end{eqnarray}
with the real notation
\begin{eqnarray}
& &  
\frac{q}{p}=d_r + i d_i,\ 
\frac{r}{p}=e_r + i e_i,\ 
\frac{1}{p}=s_r - i s_i,
\csi=c s_i.
\nonumber
\end{eqnarray}

	\item 
	The third and last example is CGL5 with one more term describing the contribution
of an intrapulse Raman scattering
\cite{NMGRK2020} 
\cite{Uzunov-et-alii2023} 
\cite{Vassilev2024},
\begin{eqnarray}
& &  {\hskip -18.0 truemm}
\hbox{(CGL5+)}\
i \GLA_t +p \GLA_{xx} +q \mod{\GLA}^2 \GLA +r \mod{\GLA}^4 \GLA + \paramb \GLA \left(\mod{\GLA}^2\right)_x -i \gamma \GLA =0,\
\label{eqCGL5Raman}
\end{eqnarray}
in which the additional parameter $\paramb$ is real.
The assumption (\ref{eqCGL35ReducMphi}) for its traveling wave reduction
also yields a third order second degree ODE for $M(\xi)$.

\end{enumerate}

\section{A privileged class of ODEs and its meromorphic solutions}
\label{sectionEremenko}

Among all the algebraic, autonomous ODEs of any order and any degree in the highest derivative,
there exists a subset of privileged ODEs,
which we call here the \textit{Eremenko class},
made of those which obey the two criteria:
\begin{enumerate}
	\item 
	The ODE possesses exactly one term
whose global degree in all the derivatives is maximal,
in short one top degree term.

	\item 
	The number of its Laurent series (excluding Taylor) is finite.
\end{enumerate}

Example:
the traveling wave reduction of the Kuramoto-Sivashinsky (KS) equation
\begin{eqnarray}
& &
u_t + \nu u_{xxxx} + b u_{xxx} + \mu u_{xx} + u u_x =0,\ \nu \not=0,
    (\nu,b,\mu) \in \mathbb{R},\
\nonumber
\end{eqnarray}
defined as
\begin{eqnarray}
& &
u(x,t)=c+U(\xi),\ \xi=x-ct,\
\nu u''' + b u'' + \mu u' + \frac{u^2}{2} + K = 0,
\label{eqKSODE}
\end{eqnarray}
in which $K$ is a real integration constant,
enjoys both properties.
Indeed, the five terms of the ODE (\ref{eqKSODE})
have the respective global degrees $1, 1, 1, 2, 0$,
i.e.~one top degree term ($u^2/2$).
The search for Laurent series 
\begin{eqnarray}
& &
u = \sum_{j=0}^{+ \infty} u_j (\xi-\xi_0)^ {j+p}, u_0 \not=0,
\nonumber
\end{eqnarray}
with $p$ a strictly negative integer and $\xi_0$ an arbitrary complex constant,
yields the unique leading term,
\begin{eqnarray}
& &
\left[p-3=2 p, \nu p(p-1)(p-2) u_0 +u_0^2/2=0\right] \Rightarrow p=-3, u_0= 120 \nu,
\nonumber
\end{eqnarray}
and none of the next $u_j$'s is arbitrary,
so the number of Laurent series is just one.                     

The reason why such ODEs are privileged is a theorem due to Eremenko,
allowing one to obtain explicitly \textbf{all} its particular solutions
whose singularities, in the complex plane of course, are only poles
(in short, meromorphic on $\mathbb{C}$).

\textbf{Theorem}
(Eremenko \cite{EremenkoKS}).
If an algebraic autonomous
ODE enjoys the above mentioned two properties,
then any solution meromorphic on $\mathbb{C}$ 
is necessarily elliptic or degenerate elliptic
(i.e.~rational in one exponential $e^{k x}$ or rational in $x$).
\label{Theorem-Eremenko}

In itself, this theorem is not constructive,
but classical, 19-th century results which we now recall make it constructive.

\section{Constructive methods implementing the theorem of Eremenko}
\label{sectionMethods}

Let us denote $N$ the (finite) number of distinct Laurent series of the ODE under consideration,
and $\Npoles$ the total number of poles, counting multiplicity.
Example: an ODE having one series with a triple pole and two series with a simple pole 
yields $N=3$ and $\Npoles=3+1+1=5$.

The constructive methods rely on the following classical results.

\begin{enumerate}
\item 
The characterization, by Briot and Bouquet \cite{BriotBouquet},
of any elliptic function $u(x)$ of elliptic order $\Npoles$ (number of poles in one period, counting multiplicity)
by a first order polynomial autonomous ODE $F(u',u)=0$
whose degrees in $u'$ and $u$ are known:
the degree of $F$ in $u'$ is the elliptic order of $u$,
and
the degree of $F$ in $u$ is the elliptic order of $u'$.
	
\item 
The generalization, by Hermite \cite{Hermite-sum-zeta}, 
to elliptic functions and their degeneracies
of the well known partial fraction decomposition of a rational function 
as the sum of a polar part 
(the sum of all negative powers near all poles)
and an entire part (a polynomial)
See details in \cite[Appendix C]{CMBook2}.
	
\item 
The construction of a method (subequation method) \cite{MC2003,CM2009} to find a closed form expression 
of all elliptic and degenerate elliptic solutions of any algebraic ODE.
The number of cases to examine is then finite for the Eremenko class of ODEs.
		
\end{enumerate}

\textit{Remark}.
The first condition required in the theorem of Eremenko can be lowered to 
``The sum of the coefficients of the top degree terms is nonzero''.
Then, together with the second condition,
the number of cases to examine is still finite, see an example in \cite{CNW-boundary-layer}.

\section{Example fourth order dispersion. A new solution}
\label{section-Application-Quartic}

In \cite{NLS-fourth-order-dispersion1994},
the authors found a pulse solution of (\ref{eqQuartic-ODE4}),
\begin{eqnarray}
& &  u(x)= a k^2 \sech^2 (k x) , a= \pm \sqrt{-\frac{5 b_4}{\gamma}},
k^2=\frac{b_2}{5 b_4^3}, \omega=\frac{24 b_2^2}{25 b_4}. 
\label{eqQuartic-pulse}
\end{eqnarray}
In order to examine whether a more general solution exists,
let us follow the successive steps of the subequation method. 

\textit{Step 1}. Find the singularity structure of the fourth order ODE (\ref{eqQuartic-ODE4}),
following for instance the guidelines in \cite{CMBook2}.
The result is:
this ODE admits two movable double poles (we omit the arbitrary origin $x_0$ of $x$)
\begin{eqnarray}
%
& & u = x^{-2} \left[a - \frac{b_2}{a \gamma} x^2 -\left( \frac{b_2^2}{a^3 \gamma^2} + \frac{\omega}{3 a \gamma}\right) x^4 
  + \left( \frac{10 b_2^3}{7 a^5 \gamma^3} + \frac{5 \omega b_2}{21 a^3 \gamma^2}\right) x^6 + K x^8 + O(x^{10}) \right],
\label{eqQuarticLaurent}
\end{eqnarray}
but an infinite number of Laurent series since an arbitrary coefficient $K$ enters the series at the index $j=8$ (a Fuchs index).

\textit{Step 2}. If possible, get rid of this positive integer Fuchs index,
by searching for a first integral as a differential polynomial of singularity degree $8$.
Such a first integral does exist here,
\begin{eqnarray}
 & & \frac{b_4}{24} \left[u''' - {u''}^2 /2\right] - \frac{b_2}{4} {u'}^2 + \gamma u^4/4 + \omega u^2/2 =K,
\label{eqQuartic-ODE3}
\end{eqnarray}
and this third order ODE now fits all Eremenko's assumptions:
autonomous,
algebraic,
one top-degree term ($\gamma u^4/4$),
finite number (two) of Laurent series since $K$ is a fixed parameter of the third order ODE.

Conclusion: \textbf{all} the meromorphic solutions of (\ref{eqQuartic-ODE3}) are elliptic or degenerate elliptic,
and,
depending on whether they possess one double pole or two, they are 
characterized by the two Briot-Bouquet subequations,
\begin{eqnarray}
& & {\hskip -9.0truemm} 
F_1 \equiv {u'}^2 + (a_{10} + a_{11} u + 0 u^2) u'+(a_{00} +a_{01} u +a_{02} u^2) - (4/a) u^3 =0,
\nonumber \\
& & {\hskip -9.0truemm} 
F_2 \equiv {u'}^4 + (a_{10} + a_{11} u + 0 u^2) {u'}^3 + \dots - (4/a)^2 u^6 =0,
\nonumber
\end{eqnarray}
whose coefficients are determined in the next steps.
What should be emphasized here is the \textit{linear} nature of the system of equations
allowing one to compute these coefficients,
making it quite easy to solve.

\textit{Step 3}.
Compute enough terms (10 is sufficient) of the two Laurent series (\ref{eqQuarticLaurent}).

\textit{Step 4}, assuming two double poles.
Require both series (\ref{eqQuarticLaurent}) to obey $F_2=0$.
This has no solution.

\begin{figure}[ht]
\begin{center}
 \includegraphics[scale=0.3]            {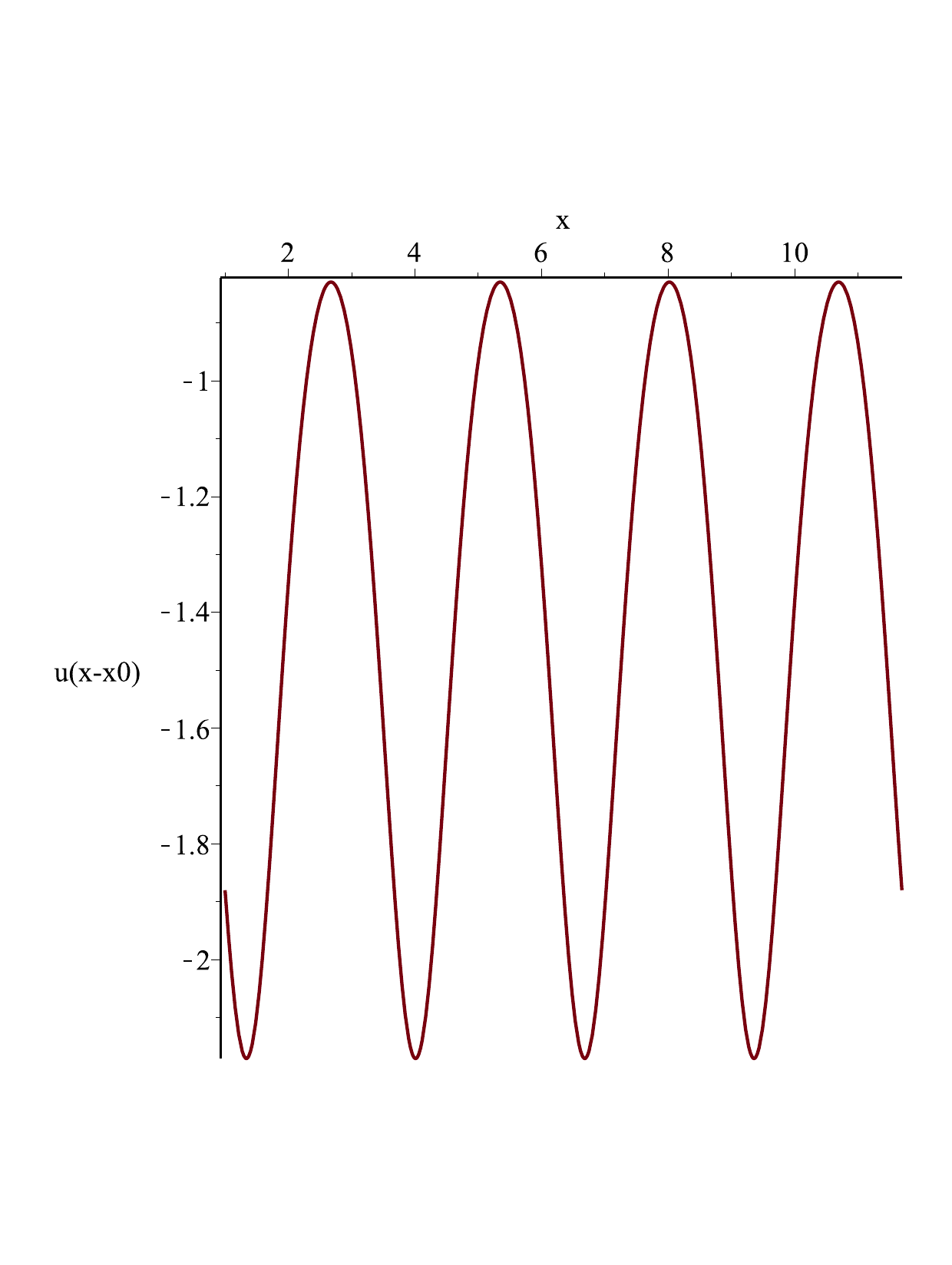}
\end{center}
\caption[Raman scattering.]
        {Raman scattering, periodic solution $u(x-x_0)$, 
$b_4=-1, b_2=1, \gamma=5, \omega=-21/5, a=1, g_2=24/5, g_3=-4/5$.
The shift $x_0$ is a half-period of $\wp$: 
 $x_0=1.337-1.198 i$, and}
the real period is $2 \Re(x_0)$. 
\label{Fig-Raman}
\end{figure}

\textit{Step 4}, assuming one double pole.
Require anyone of the two series (\ref{eqQuarticLaurent}) to obey $F_1=0$.
The result is one and only one solution $F_1=0$,

\begin{eqnarray}
& &
\left\lbrace
\begin{array}{ll}
\displaystyle{
F_1 \equiv 3 a^4 \gamma^3 {u'}^2 
-12 a^3 \gamma^3 u^3 
+36 a^2 b_2 \gamma^2 u^2
-(20 a^3 \gamma^2 \omega + 96 a b_2^2 \gamma) u
+40 a^2 b_2 \gamma \omega +192 b_2^3=0,
}\\ \displaystyle{
\gamma K=\frac{5}{108}    \left(\omega - \frac{48 b_2^2}{5 b_4} \right)\left(\omega - \frac{24 b_2^2}{25 b_4} \right),
}
\end{array}
\right.
\nonumber
\end{eqnarray}
This is an affine transform of the canonical equation of Weierstrass,
with the general solution
\begin{eqnarray}
& &
u=a \left(\wp(x,g_2,g_3)+ \frac{5 b_4}{\gamma}\right),
g_2=\frac{4}{3 b_4}        \left(\omega - \frac{3 b_2^2}{5 b_4}\right),
g_3=\frac{4 b_2}{15 b_4^2} \left(\omega - \frac{6 b_2^2}{5 b_4}\right),
\nonumber
\end{eqnarray}
it is bounded for some set of parameters (see, e.g., Figure \ref{Fig-Raman})
and therefore physically admissible.
This elliptic solution reduces to the pulse solution (\ref{eqQuartic-pulse})
for the value $\omega=(24/25) b_2^2/b_4$.

\textit{Remark 1}.
Since the discriminant $g_2^3-27 g_3^2$ vanishes for two other values of $\omega$,
there could exist two other pulse solutions on a nonzero background
$u=a k^2 \sech^2 (k x) + c_0 k^2$, $c_0 \not=0$,
but their values of $\omega$ are not real.

\textit{Remark 2}.
The invariance of (\ref{eqQuartic-ODE3}) under $u \to -u$ suggests to process
the ODE for $u^2$,
since it also obeys the conditions of Eremenko 
(one top degree term $\gamma (u^2)^5/4$, one Laurent series with a quadruple pole),
we leave that to the interested reader.
This could provide new solutions $u(x)$ as the square roots of elliptic functions.


\section{Example CGL5. A nondegenerate elliptic solution}
\label{section-Application-CGL5}

The third order ODE (\ref{eqCGL35Order3}) for $M(\xi)$ 
admits for $e_i \not=0$ (CGL5 case) 
exactly one top degree term $-4 e_r e_i^2 M^{10}$
and
exactly four Laurent series
\cite[Eq.~(21)]{ConteNgCGL5_ACAP}
\begin{eqnarray}
& & {\hskip -15.0 truemm}
M=A_0^2 \xi^{-1}
\left[
1+\left(\frac{\csi}{4}+\frac{2 d_r A_0^2-2 e_i d_i A_0^6} {4(1+4\alpha^2)}\right) \xi
+\mathcal{O}(\xi^2)
\right],
\label{eqCGL5-M-Laurent-series}
\end{eqnarray}
in which the pair $(A_0^2,\alpha)$ of real constants takes four values \cite{MCC1994}, 
\begin{eqnarray}
& & {\hskip -13.0 truemm}
\hbox{(CGL5) }
\left(-\frac{1}{2}+i \alpha\right) \left(-\frac{3}{2}+i \alpha\right) p + A_0^4 r=0,\
\alpha^2 -2 \frac{e_r}{e_i} \alpha -\frac{3}{4}=0,
A_0^4=\frac{2 \alpha}{e_i}, e_i \not=0.
\label{eqCGL5LeadingOrderComplex}
\end{eqnarray}
The restriction $e_r\not=0$ can be removed \cite{ConteNgCGL5_ACAP},
and the conclusion of Eremenko (``meromorphic implies elliptic'')
holds  
for all values of the CGL5 parameters $p, q, r$ (complex), $\gamma$ (real)
and of the traveling waves parameters $c, \omega$ (real).

Let us exemplify here the search for nondegenerate elliptic solutions.
Such solutions are easier to find for two reasons.

The first reason is the necessary condition of the vanishing, inside a period parallelogram,
of the sum of the residues of the considered Laurent series (\ref{eqCGL5-M-Laurent-series}) of $M$ 
(or more generally of any rational function of $M$ and its derivatives).
As done in \cite[\S 3.1]{ConteNgCGL5_TMP},
the number of series involved in this sum must be equal to four,        
the number of terms in each series must be at least equal to seven,
and in the generic case $\csi$ arbitrary
the four monomials $M^2$, $M^3$, $M^4$, ${M'}^2$ are enough 
to generate, \textit{via} the necessary conditions sum(residues(monomial))=0,
the constraints \cite[Eq.~(25)]{ConteNgCGL5_TMP},
\begin{eqnarray}
& & 
\forall \csi:\
(M^2 \Rightarrow e_r=0),
(M^3 \Rightarrow d_r=0),
(M^4 \Rightarrow 16 g_i + 3 \csi^2=0),
({M'}^2 \Rightarrow d_i=0).
\nonumber
\end{eqnarray} 

The second reason is a simplification in the first order subequation for $M$,
characterized by four simple poles,
\begin{eqnarray}
& & {\hskip -9.0truemm} 
 F \equiv \sum_{k=0}^{4} \sum_{j=0}^{8 -2k} a_{j,k} M^j {M'}^k=0,\ a_{0,4}\not=0.
\label{eqsubeqODEOrderOnePP}
\end{eqnarray}
Indeed, a not so well known result of Briot and Bouquet \cite[\S 181 p.~278]{BriotBouquet}
is that,
in order for this first order ODE $F=0$ to have a nondegenerate elliptic general solution,
it should not contain the power one of $M'$, 
thus canceling the seven coefficients corresponding to $k=1$ in (\ref{eqsubeqODEOrderOnePP}).

The explicit expression of $F$, Eq.~(\ref{eqsubeqODEOrderOnePP}),
can be found in \cite[Eq (47)]{ConteNgCGL5_TMP}.
In order to present the methods of its integration,
let us consider its particlular case $\csi=0$,
in which $F$ reduces to an equation first isolated by Vernov \cite{VernovCGL5},
\begin{eqnarray}
& & q=0, e_r=0, g_i=0, \csi=0:\
F \equiv e_i (3 M')^4 - M^2 \left(3 e_i M^2 -4 g_r\right)^3=0.
\nonumber
\end{eqnarray}
\medskip

At least three methods exist to integrate this ODE.

\begin{enumerate}
	\item The first one is to notice its binomial type ${M'}^m = \hbox{polynomial}(M)$,
a class already integrated by Briot and Bouquet \cite{BriotBouquet}.
Its solution is therefore a homographic transform of $\wp^2$,
see (\ref{eqCGL5-Vernov-solution-M}).

	\item Hermite decomposition.
	This second method is to represent the solution $M$ by its Hermite decomposition,
the sum of a constant term and four simple poles of residues the four values of $A_0^2$, 
see (\ref{eqCGL5LeadingOrderComplex}),
\begin{eqnarray}
& & 
M=c_0+A_0^2 \sum_{j=0}^3 i^j \zeta(\xi-\xi_j), \xi_0=0, \zeta'=-\wp,
\label{eq-CGL5-M-Hermite}
\end{eqnarray}
the unknowns being $c_0, \wp(\xi_j), \wp'(\xi_j)$.
The technique to compute them efficiently has been explained by Demina and Kudryashov \cite{DK2011method},
this is to identify the four Laurent series (\ref{eqCGL5-M-Laurent-series})
to the four expansions of (\ref{eq-CGL5-M-Hermite})
near $\xi=\xi_j - \xi_0$.
Finally, the identity \cite[Chap.~18, \S 18.4.3]{AbramowitzStegun}
\begin{eqnarray}
& & \forall z_1,z_2:\
\zeta(z_1+z_2)=\zeta(z_1)+\zeta(z_2)+\frac{1}{2}\frac{\wp'(z_1)-\wp'(z_2)}{\wp(z_1)-\wp(z_2)}
\nonumber
\end{eqnarray}
converts (\ref{eq-CGL5-M-Hermite}) to (\ref{eqCGL5-Vernov-solution-M}).
 
	\item The third method is to use the very nice package \verb+algcurves+ \cite{MapleAlgcurves}
of the computer algebra language Maple \cite{Maple}.
The command \verb+Weierstrassform(F,M,M',X,Y,Weierstrass)+
returns the birational transformation between the equation $F(M,M')=0$ and 
the canonical Weierstrass equation $Y^2= 4 X^3 - g_2 X - g_3$,
i.e.~four rational functions $M(X,Y)$, $M'(X,Y)$, $X(M,M')$, $Y(M,M')$.
However, because of the existence of an addition formula for $\wp$ \cite[Chap.~18, \S 18.4.1]{AbramowitzStegun},
these rational functions may be uselessly complicated,
but they only differ from (\ref{eqCGL5-Vernov-solution-M}) by a shift of $\xi$,
see such an example in \cite[Eq (45)]{ConteNgCGL5_ACAP}.

\end{enumerate}

\section{Example CGL5 + term $ \paramb A \partial_x \mod{A}^2$}
\label{section-Application-CGL5plus}

This is in fact not an example, but a suggestion to the reader
to possibly obtain new, physically interesting singlevalued solutions 
of (\ref{eqCGL5Raman}).
Indeed, the additional real parameter $\paramb$ does not alter the singularity structure of 
the third order ODE for $M$ (four simple poles)
and the method used in \cite{ConteNgCGL5_ACAP}
could probably also conclude that meromorphic solutions
are finitely many and necessarily elliptic or degenerate.

Therefore, following the guidelines of Ref.~\cite{CMNW-CGL35-Article},
it would be possible to obtain all those solutions in closed form.
One of the challenges would be to determine the values of $\paramb$, if any,
defining a nondegenerate elliptic solution bounded on the real axis.


\section{A method for nonmeromorphic exact solutions}
\label{section-nonmeromorphic}

In 2020, Nisha \textit{at alii} \cite{NMGRK2020}
(see also 
\cite{Uzunov-et-alii2023} 
\cite{Vassilev2024})
found a new closed form solution $M(\xi)$ of the ODE for the square modulus of the PDE (\ref{eqCGL5Raman})
for CGL5 + term $\paramb A \partial_x \mod{A}^2$,
by a very simple method,
which is worth being presented here.

In the ODE for $M(\xi)$ (which admits four Laurent series with a simple pole
but is generically outside the scope of Eremenko's theorem),
they do not assume $M$ to obey an ODE of the form of Briot and Bouquet
\begin{eqnarray}
& & {\hskip -9.0truemm} 
 F \equiv \sum_{k=0}^{m} \sum_{j=0}^{2m-2k} a_{j,k} u^j {u'}^k=0,\ a_{0,m}\not=0,
\nonumber
\end{eqnarray}
for some integer $m \le 4$,
as done in the case $\paramb=0$ \cite{CMNW-CGL35-Article}.
Instead of that,
they set $M=\rho^2$,
which defines a multivalued function $\rho(\xi)$,
and,
at least in the simplest situation $m=1$ of only one simple pole for $M$,
they assume $\rho$ to obey a first order, autonomous, algebraic, Abel ODE
matching the singularity structure,
\begin{eqnarray}
& & M=\rho^2,
\rho'=\frac{k}{2} \rho^3 + c_2 \rho^2 + c_1 \rho+ c_0. 
\label{eqAbel-subeq}
\end{eqnarray}
The third order ODE for $M$ then evaluates to a polynomial in $\rho$,
which is required to identically vanish.

Because of the unnecessary restriction which they impose
\begin{eqnarray}
& & \varphi'= \alpha_0 + \alpha_2 \rho^2,
\label{eqNishaRestriction}
\end{eqnarray}
they only find one new solution in which $M$ is multivalued,
characterized by the Abel subequation
\begin{eqnarray}
& & M=\rho^2,
\rho'= (\rho-a)^2 (\rho+a).
\label{eqNishi-Abel}
\end{eqnarray}
The function $\rho$ is then a homographic transform of the Lambert function $W(\xi)$ \cite{Lambert-function}
\begin{eqnarray}
& &
\rho=a \left(1-\frac{2}{1+W\left(e^{1+4 a^2 \xi}\right)} \right),
\frac{\D W}{\D z}=\frac{W}{z(1+W)},
\nonumber
\end{eqnarray}
whose general solution $W$ is multivalued in the complex plane.
Since the variable $\xi$ is real in the considered physical problem,
this solution $M(\xi)$ respresents a kink \cite[Fig.~4]{NMGRK2020},
different from the usual $\tanh(K \xi)$ kink.

If the restriction (\ref{eqNishaRestriction}) is removed, 
this method provides three Abel subequations
associated to four sets of constraints between all the parameters,
\begin{eqnarray}
& & 
                 \rho'=\frac{k}{2} \left(\rho^2 + \frac{2 c s_i}{k}\right) \left(\rho+ 2 \frac{c_2}{k}\right), k^2=2 e_r, b=\frac{e_i}{k s_i},
\nonumber \\ & & \rho'=\frac{k}{2} \left(\rho^2 + \frac{2 c s_i}{3 k}+ 2 \frac{c_2}{k} \rho\right) \rho,       k^2=2 e_r,
\nonumber \\ & & \rho'=\frac{k}{2} \left(\rho^3 + \frac{2 c s_i}{  k} \rho+  \frac{2 c_0}{k}    \right),            e_r=0, b=\frac{e_i}{k s_i}.
\nonumber
\end{eqnarray}
These Abel equations cannot be linked anymore to the Lambert function,
but their (multivalued) solution can be parametrized as follows,
\begin{eqnarray}
& & 
\rho=t, M=t^2, \xi=\xi_0+\int \frac{\D t}{\frac{k}{2} t^3 + c_2 t^2 + c_1 t+ c_0}\ccomma
\nonumber
\end{eqnarray}
for instance in the case of (\ref{eqNishi-Abel}),
\begin{eqnarray}
& & 
\rho=a t, M=a^2 t^2, \xi=\xi_0- \frac{1}{2 a^2 (t+1)} + \frac{1}{4 a^2} \log\frac{t+1}{t-1}\cdot
\nonumber
\end{eqnarray}

\textit{Remark}.
Assumptions more general than (\ref{eqAbel-subeq}) could yield additional solutions,
provided of course that they respect the singularity structure.

\section*{Acknowledgements}

We thank
Alejandro Aceves for bringing our attention to Ref \cite{NLS-fourth-order-dispersion1994}.
RC is pleased to thank the Institute for Mathematical Research of The University of Hong Kong,
and the Institute of Advanced Study of Shenzhen university
for their generous support.
NTW was partially supported by the RGC grant 17307420. 
WCF was supported by the National Natural Science Foundation of China (grant no.~11701382).


\vfill\eject

\end{document}